\documentstyle[twoside,fleqn,espcrc2,graphicx,epsfig,xspace,amssymb,amsmath]{article}

\title{Perfect Gauge Actions on Anisotropic Lattices\thanks{Work supported in part by Schweizerischer Nationalfonds}
\thanks{Presented by P. R\"ufenacht}}

\author{Philipp R\"ufenacht\address{Institute for Theoretical Physics,
University of Bern, Sidlerstrasse 5, CH-3012 Bern, Switzerland}, 
Urs Wenger\address{Theoretical Physics, Oxford University, 1 Keble Road, Oxford OX1 3NP, United Kingdom}}

\begin{document}
\setlength{\mathindent}{0pt}
\begin{abstract}
On anisotropic lattices, where generally the
lattice is rather coarse in spatial directions, a parametrized classically perfect action
could help reducing lattice artifacts considerably.
We investigate the possibility of constructing such actions for SU(3) gauge theory.
We present two different methods to do so, either repeating the procedure used to
create our newly parametrized isotropic FP action, or performing one single step
starting with the isotropic result.
The anisotropic action is parametrized using an ansatz including anisotropically
APE-like smeared (``fat'') links. The parametrized classically perfect action
with anisotropy $\xi=a_s/a_t=2$
is constructed and  the renormalized anisotropy is measured using the
torelon dispersion relation. It turns out that the renormalization is small.

\end{abstract}

\maketitle

\section{INTRODUCTION}
Measuring heavy states (e.g.~glueballs) on mo\-de\-rately large lattices is
feasible using anisotropic lattices whose resolution in temporal direction
is finer than the one in spatial directions \cite{Morningstar:1997ff,Morningstar:1999rf,Juge:1998aa,Manke:1999aa,Drummond:1999aa,Manke:1999bb,Liu:2000ce}. Having rather coarse (spatial)
lattices calls for the use of improved actions; a radical choice there is
the classically perfect FP action \cite{Hasenfratz:1994sp} which has been constructed for (isotropic)
SU(3) gauge theory \cite{DeGrand:1995ji,DeGrand:1995jk,DeGrand:1996ab,Blatter:1996ti}.
Our new parametrization \cite{Niedermayer:2000yx,Niedermayer:2000tp} can serve as a
basis for the construction of the corresponding anisotropic action as will
be shown later. Anisotropic classically perfect gauge actions are an independent approach
for measuring heavy states in pure gauge theory and thus probe universality.

\section{CREATING PERFECT ACTIONS}
In this paper we consider SU(N) pure gauge theory in $d=4$ Euclidean
space on a periodic lattice, the numerical work is done for SU(3).

For asymptotically free theories, the classically perfect FP action is
determined by a saddle point equation \cite{Hasenfratz:1994sp}:
\begin{equation}
  {\cal A}^{\text{FP}}(V) = \min_{\{U\}} \left\{{\cal A}^{\text{FP}}(U) + T(U,V)\right\},
\label{saddlepoint}
\end{equation}
where $T(U,V)$ is the blocking kernel that defines the block transformation connecting
the initial (fine) configuration $U$ and the blocked (coarse) configuration $V$.

In order to get a FP action for coarse configurations occuring in MC simulations,
eq.~(\ref{saddlepoint}) is used as a recursion relation connecting an action on a coarse
configuration $V$ to one on a finer configuration $U$. The recursion step is
iterated $3\sim 4$ times using some appropriate simple starting action ${\cal A}(U)$
on the smoothest configuration.
In principle the final FP
action contains infinitely many couplings and thus has to be parametrized such
that it may be used in a numerical simulation.

\section{GETTING ANISOTROPIC}
We pursue two different possibilities of modifying the method sketched in the previous section
to obtain an \emph{anisotropic} action. Firstly, the procedure may be
repeated with an anisotropic starting action for smooth configurations, which may
be a simple anisotropic discretization of the continuum action. Due to the isotropic
nature of the blocking the anisotropy is relayed up to the level of coarse
configurations.

Secondly, we can start with the
parametrized isotropic FP action on coarse configurations and perform one or several
additional \emph{purely spatial} blocking steps each doubling the anisotropy $\xi=a_s/a_t$
(if one is working with scale-2 blocking steps) such that one obtains actions with anisotropies
$\xi$=2, 4, 8, $\ldots$. These actions are not \emph{Fixed Point} actions in the
narrower sense as the action on the l.h.s. of eq.~(\ref{saddlepoint}) has got an
anisotropy twice as large as the one on the r.h.s. However, the actions are
still classically perfect. In particular, they possess exactly scale invariant
lattice instanton solutions.

\section{SCALAR FIELDS, PERTURBATIVE TESTS}
For scalar fields it can be proven analytically that both methods mentioned
in the last section are completely equivalent and yield the same result.
Furthermore it can be shown that the couplings for anisotropic actions on
scalar fields can be made short-ranged in all directions, this has already been
studied earlier in \cite{Bietenholz:1999aa}.

\begin{figure}[htbp]
\begin{center}
\includegraphics[width=5.3cm,angle=-90]{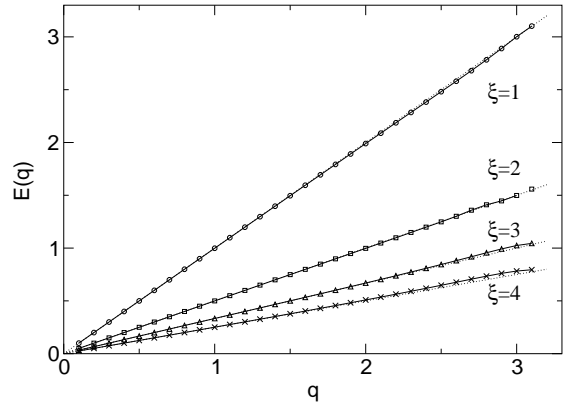}
\end{center}
\vspace{-0.8cm}
\caption{Spectra for different anisotropies, quadratic approximation to the
perfect action. The dotted lines represent the continuum result.}
\label{quadspec}
\end{figure}

In addition, we study gauge fields in the quadratic approximation, where the action
is
\begin{equation}
{\cal A}=\frac{1}{2N_c}\sum_{n,r} \rho_{\mu\nu}(r)\text{Tr}[A_\mu(n+r)A_\nu(n)]
\end{equation}
neglecting terms of order 3 and higher in the gauge potential.
Fig.~\ref{quadspec} displays the nice spectra obtained for different
anisotropies. Also, in the perturbative potential the anisotropic classically perfect 
actions perform as well as the isotropic FP action.

\section{ANISOTROPIC STARTING ACTION}
Repeating the isotropic FP action program \cite{Niedermayer:2000yx,Niedermayer:2000tp}
with an anisotropic starting action has got the advantage that the recursion steps are
exactly the same as for the isotropic action, i.e.~exactly the same
RG transformation may be used. Furthermore, one may construct perfect actions of any desired anisotropy $\xi$.
However, the procedure is costly as one has to perform
again several RGT and fitting steps for each value of $\xi$. Besides, in each step it has to be
carefully checked that the action parametrizations used are accurate on 
the respective fluctuations and that the renormalization of the anisotropy is small.
That is why we prefer to perform a small number of 
purely spatial blocking steps starting with the isotropic action parametrized
in \cite{Niedermayer:2000yx,Niedermayer:2000tp}.

\section{SPATIAL BLOCKING STEP}
To perform a purely spatial blocking step we modify the symmetric blocking kernel \cite{DeGrand:1995ji}:
\begin{multline}
  T(U,V) =\\
 - \frac{\kappa}{N_c} \sum_{n_B,\mu} \left(\text{Re} \text{Tr}[V_\mu(n_B)
  Q^\dagger_\mu(n_B)] - {\cal N}_\mu^\beta \right),
\end{multline}
where $Q_\mu(n_B)$ is a mean of products of fine links that
connect two coarse lattice sites (we use $Q_\mu$ of type III blocking, defined in \cite{Blatter:1996ti}).
Thus, the fine links are first smeared and subsequently blocked to the coarse lattice.
The normalization term ${\cal N}_\mu^\beta$ assures that
the partition function remains unchanged under the corresponding RG 
transformation. The parameter $\kappa$ determines the weight of
the blocking kernel relative to the action value ${\cal A}(U)$ and thus the freedom of the fine link products to fluctuate with
respect to the coarse links.

We split the sum in $T(U,V)$ into a spatial and a temporal part with different
$Q_\mu(n_B)$ and values of $\kappa$.
Whereas $Q_i(n_B)$ is kept the same for the spatial directions $i=1,2,3$,
$Q_4(n_B)$ now connects two coarse lattice sites whose (temporal) lattice
distance is the same as on the initial fine lattice, i.e.~the links are only
smeared but not blocked.
The parameters $\kappa_t$ and $\kappa_s$ are independent of
each other and can be chosen differently to make the resulting action
accurately parametrizable with a small number of parameters.

\section{ANISOTROPIC PARAMETRIZATION}
To parametrize the anisotropic actions we generalize our
``fat link'' parametrization described in \cite{Niedermayer:2000yx}:
\begin{equation}
{\cal A}[U]=\frac{1}{N_c}\sum_x\sum_{\mu<\nu}\sum_{k,l} p_{kl}\,u_{\mu\nu}(x)^k\, w_{\mu\nu}(x)^l\,,
\end{equation}
\begin{eqnarray}
u_{\mu\nu} & = & \text{Re Tr}(1-U_{\mu\nu}^{\text{pl}}),\nonumber\\
w_{\mu\nu} & = & \text{Re Tr}(1-W_{\mu\nu}^{\text{pl}}),
\end{eqnarray}
where $U_{\mu\nu}^{\text{pl}}$, $W_{\mu\nu}^{\text{pl}}$ denote the pla\-quettes built of
standard and smeared gauge links, respectively. 

Firstly, the coefficients $p_{kl}$ are chosen differently for pla\-quette orientations
parallel ($p_{kl}^{\text{tm}}$) and perpendicular ($p_{kl}^{\text{sp}}$) to the temporal direction.
Secondly, the construction of the smeared pla\-quettes is altered:
There are three different kinds of smeared links contributing to a smeared
pla\-quette $W_{\mu\nu}^{\text{pl}}$: 
\begin{center}
\includegraphics[width=1.2cm]{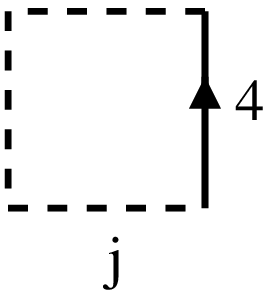}\hspace{1cm}
\includegraphics[width=1.5cm]{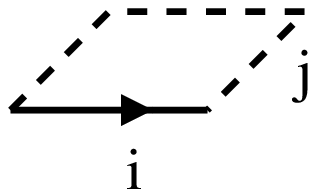}\hspace{1cm}
\includegraphics[width=1.2cm]{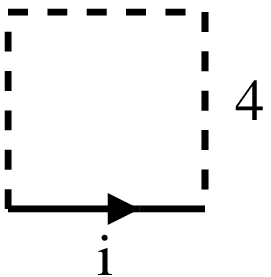}
\end{center}
temporal links (always contributing to
a temporal pla\-quette), spatial links that contribute to a spatial pla\-quette
and spatial links that contribute to a temporal pla\-quette.

In addition, there are four kinds of staples contributing to a smeared link which are ``asymmetric'' either
because they lie in the same plane as the pla\-quette that is going to be built or because they
go in time direction:
\begin{center}
\includegraphics[width=2cm]{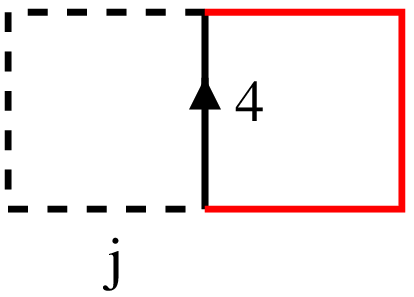}
\includegraphics[width=1.5cm]{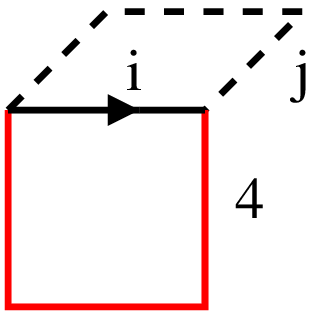}
\includegraphics[width=2cm]{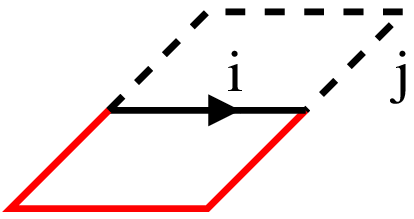}
\includegraphics[width=1.2cm]{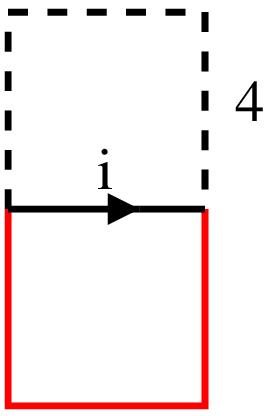}
\end{center}

Because of this we introduce four different asymmetry parameters $\eta$
(weights of the staples) and for each iteration level $i$ three different
parameters $c_i$ (specifying the expansion of the smeared links) \cite{Niedermayer:2000yx}. These additional parameters do
not influence the overhead of the action significantly, however the time used
for parametrizing increases considerably.

Tests with reduced sets of actions (smaller number of non-zero $\eta$
parameters or only one $c_i$ value for all pla\-quette orientations) show that in order
to fit well the perfect action the sophisticated parametrization is indeed
needed.

\section{THE $\mathbf{\xi=2}$ ACTION}

The quadratic approximation suggests a ratio of $\kappa_t/\kappa_s=\xi^2$ for the
$\kappa$-parameters in the anisotropic blocking kernel. As the blocking is unchanged
in spatial direction, $\kappa_s$ is kept at the optimal value of $\kappa_s=8.8$ obtained for the isotropic
case; thus $\kappa_t=35.2$. Fits on 40 configurations are performed and yield an
action with 3 levels of $c_i$, $0<k+l\le 4$ for $p_{kl}^{\text{sp}}$ and $0<k+l\le 3$
for $p_{kl}^{\text{tm}}$. The fit is stable as the result does not change considerably when adding 
further independent configurations, and extending the parameter set by increasing
the number of $c$-levels or the number of linear parameters does not further decrease $\chi^2$ 
significantly.

\section{THE RENORMALIZATION OF THE ANISOTROPY}
\begin{figure}[htbp]
\begin{center}
\includegraphics[width=5.3cm,angle=-90]{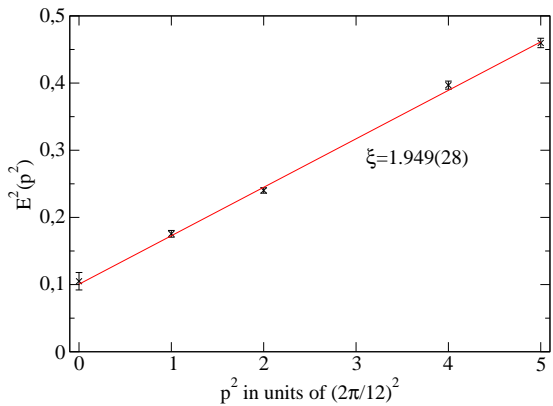}
\end{center}
\vspace{-0.8cm}
\caption{Torelon dispersion relation. The straight line is the correlated fit to $E^2(p)=m_T^2+p^2$ in
the range $p^2=1..4$.}
\label{tordisp}
\end{figure}

The anisotropy of the above action is measured using the torelon dispersion relation as
described in \cite{Teper:1999aa,Alford:2000an}. On a $12^2\times 6\times 30$ lattice at $\beta=3.3$ with
periodic boundary conditions we measure the Polyakov line around the short spatial direction
on APE-smeared configurations of smearing levels 3, 6, 9, 12, 15 with smearing parameter $\lambda_s=0.1$
(see \cite{Niedermayer:2000yx}). 23600 alternating Metropolis and over-relaxation sweeps and 4720
measurements are performed. A fully correlated fit using variational techniques yields for the
renormalized anisotropy $\xi_R=1.949(28)$; the renormalization of the input anisotropy $\xi_0=2$ is
thus very small. A rough estimate of the scale yields $a_s=0.153(8)$~fm and thus $a_t=0.079(5)$~fm.
Fig.~\ref{tordisp} shows the lower part of the dispersion relation including the result of the
correlated fit in the range $p^2=1..4$. Taking into account larger fitting ranges up to $p^2=0..8$, we obtain
results that are all consistent with the quoted value of $\xi_R$ within statistical errors.

\section{CONCLUSIONS AND PROSPECTS}
The $\xi=2$ classically perfect gauge action has been constructed using one spatial blocking step
starting with the isotropic FP action.
It shows a very small renormalization of the anisotropy at $\beta=3.3$ corresponding
to $a_s\approx 0.15$~fm. Scaling tests will be performed on this action measuring the
static quark--antiquark potential (including off--axis separations to probe rotational
invariance) and glueball masses. Furthermore, the spatial blocking procedure sketched above will be repeated
using the $\xi=2$ parametrized classically perfect action to build a perfect $\xi=4$ action.

{\bf Acknowledgements:} We would like to thank Ferenc Niedermayer for useful suggestions and discussions.

\end{document}